# Nanophotonic Extreme Ultraviolet Polarimeter

Anna Kulter[1], Lorenz Weiß[1], David J. Grafinger[1], Alexander Sagar Grossek[1], Rodolfo Previdi[2], Martin Schultze[1], Marcus Ossiander[1, 3, *]

1 Institute of Experimental Physics, Graz University of Technology, Austria.
2 Institute of Science and Technology Austria, Austria.
3 John A. Paulson School of Engineering and Applied Sciences, Harvard University, USA.
* Corresponding author: marcus.ossiander@tugraz.at

**Nanophotonic engineering has revolutionized how we manipulate the polarization of visible light. Here, we extend such control to extreme ultraviolet radiation, the spectral region of attosecond science and semiconductor lithography, where transmissive optical elements are lacking. We implement channel waveguides that restrict incident 41-nm-wavelength light to a few modes with defined spatial profiles and propagation constants. By designing the propagation-length-dependent interference of two eigenmodes, we achieve polarization-dependent forward scattering, which we exploit to create a transmissive nanophotonic analyzer that resolves the polarization axis and spectrum of extreme ultraviolet light from a single measurement. With a 0.9°-sensitivity for polarization rotations in a 1-second measurement and single-shot capability, the device opens new experimental opportunities in ultrafast magneto-optic spectroscopy and at accelerator-based light sources.**

A photon's frequency and polarization determine how it interacts with matter. On the one hand, this fact renders tracking the polarization-dependent spectrum of absorbed or reflected extreme ultraviolet (EUV) light a sensitive probe of the fastest dynamics in matter and has enabled, e.g., investigating ultrafast magnetism (*1*–*6*), molecular chirality (*7*, *8*), and valley dynamics (*9*). On the other hand, the polarization of light introduces engineering constraints to high-numerical-aperture semiconductor lithography and, if not controlled, can diminish resolution (*10*). Consequently, there is a growing need for analyzing and controlling the polarization state of EUV light.

The scarcity of transmissive optics still renders resolving the polarization of EUV light challenging. Current approaches mainly rely on precise multi-mirror assemblies (*11*), mechanically moving polarimeters (*6*, *12*, *13*) or dichroism measurements (*8*, *14*–*16*), although grating-based reflective polarization analyzers (*17*) and absorption-based polarizers for the EUV spectrum were theoretically proposed (*18*) and experimentally realized (*19–21*).

Due to their thinness, metasurfaces circumvent the universal material absorption in the EUV spectrum and allow the realization of phase optics in this scientifically and technologically important spectral region (*22–27*). However, controlling small-wavelength light with low-refractive-index materials requires minuscule feature sizes and large aspect ratios, such that nanofabrication constraints so far prevent applying the wavelength-scale polarization-shaping capabilities of oblong nanostructure metasurfaces (*28–30*) to EUV light.

Here, we provide an experimentally viable solution: we demonstrate that multimode interference in photonic waveguides is designable even in the EUV spectrum, enables precise and polarization-dependent control of the spatial light profile at their end facet, and allows creating a transmissive EUV analyzer that emulates the function of a partially polarizing Wollaston prism and simultaneously provides spectral resolution.

EUV light scattering from a nanovoid: We first examine EUV light scattering from a single cylindrical void (diameter d = 0.14 μm, length h = 0.27 μm) encased in crystalline silicon. An incident x-polarized plane wave propagates along the z-direction and hits the cylinder's base normally (Fig. 1A). In silicon, EUV light with photon energy $\hbar\omega = 30$ eV (vacuum wavelength $\lambda_0 = 41.3$ nm, vacuum wave vector $\mathbf{k} = (0, 0, k_0)^T = (0, 0, 152.0\ \mu m^{-1})^T$) experiences a complex refractive index $\tilde{n}_{Si} = n_{Si} + i\kappa_{Si} = 0.860 + 0.014i$, leading to an attenuation and an accelerated phase velocity compared to light traveling in vacuum ($\tilde{n}_{vacuum} = 1$). In turn, light scatters at the void; Fig. 1B shows the resulting scattered field (finite-difference-time-domain modeling, tidy3d, see supplementary materials). Whereas the scattered field distributions in the x-z (transverse magnetic, TM) plane and in the y-z (transverse electric, TE) plane appear similar, subtracting the magnitudes of the two field distributions (Fig. 1C) reveals direction-dependent forward scattering by the cylinder: for a scattering angle of 14.7° (dashed lines in Fig. 1C), EUV light is more likely to scatter in the TM-plane than in the TE-plane, i.e., it is more likely to gain transverse momentum along the x-axis ($k_x = \pm n_{Si}\ k_0 \sin(14.7°) = \pm 33.2\ \mu m^{-1}$) than along the y-axis ($k_y = \pm 33.2\ \mu m^{-1}$). Doubling the channel's length (h = 0.54 μm, Fig. 1D) reverses the situation (preferentially redirecting light in the TE-plane), suggesting an accumulating propagation phase as the cause of the direction-dependent forward scattering. Furthermore, because the system is circularly symmetric around the optical axis, the direction-dependent forward scattering must originate from the incident light's polarization and enable polarization analysis.

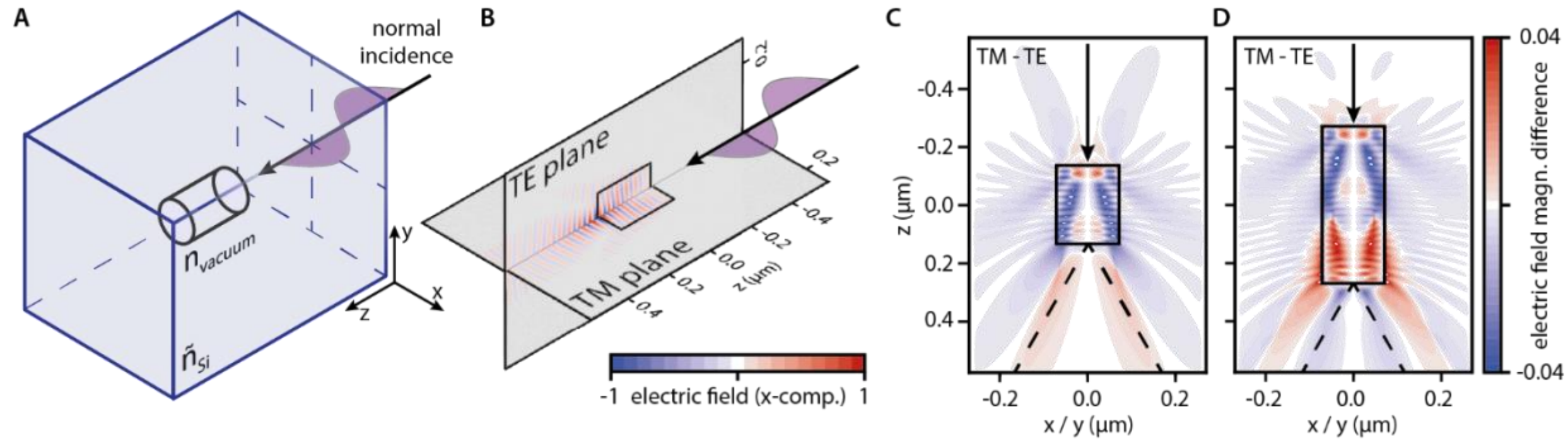

**Fig. 1. Extreme ultraviolet (EUV) forward scattering by a void in silicon.** (**A**) Simulation setup: an x-polarized plane wave illuminates a 0.27-µm-long vacuum cylinder encapsulated by silicon. (**B**) x-component of the scattered electric field in the x-z- and y-z-planes. The black rectangles mark the cylinder. (**C**) Difference between the scattered electric field magnitudes in the two planes shown in (**B**) (color bar in (**D**)). (**D**) Same as (**C**) but for a 0.54-µm-long cylinder.

EUV few-mode waveguides: To describe this behavior, we treat the vacuum channel (high-refractive-index) in silicon (low-refractive-index) as an optical fiber and examine its eigenmodes (*22*, *31*, *32*). A refractive index contrast of $n_{vacuum} - n_{Si} = 0.14$ yields a high numerical aperture $\mathrm{NA} = \sqrt{(n_{\mathrm{vacuum}}^2 - n_{\mathrm{Si}}^2)} = 0.51$ and indicates strong guiding (*33*). Only two of the 16 supported eigenmodes, $HE_{11}$ and $HE_{12}$, are excited by an x-polarized plane wave incident along the cylinder axis (see supplementary materials, Tab. S1, Fig. S1, and Fig. S2). Figs. 2A, B show their transverse electric field profiles at the position of excitation. Both modes are predominantly confined to the vacuum, yielding low effective extinction coefficients ($\kappa^{\mathrm{HE}_{11}} = 3.6 \times 10^{-4}$, $\kappa^{\mathrm{HE}_{12}} = 3.0 \times 10^{-3}$), allowing them to propagate 39 ($HE_{11}$) / 5 ($HE_{12}$) times farther than in bulk silicon.

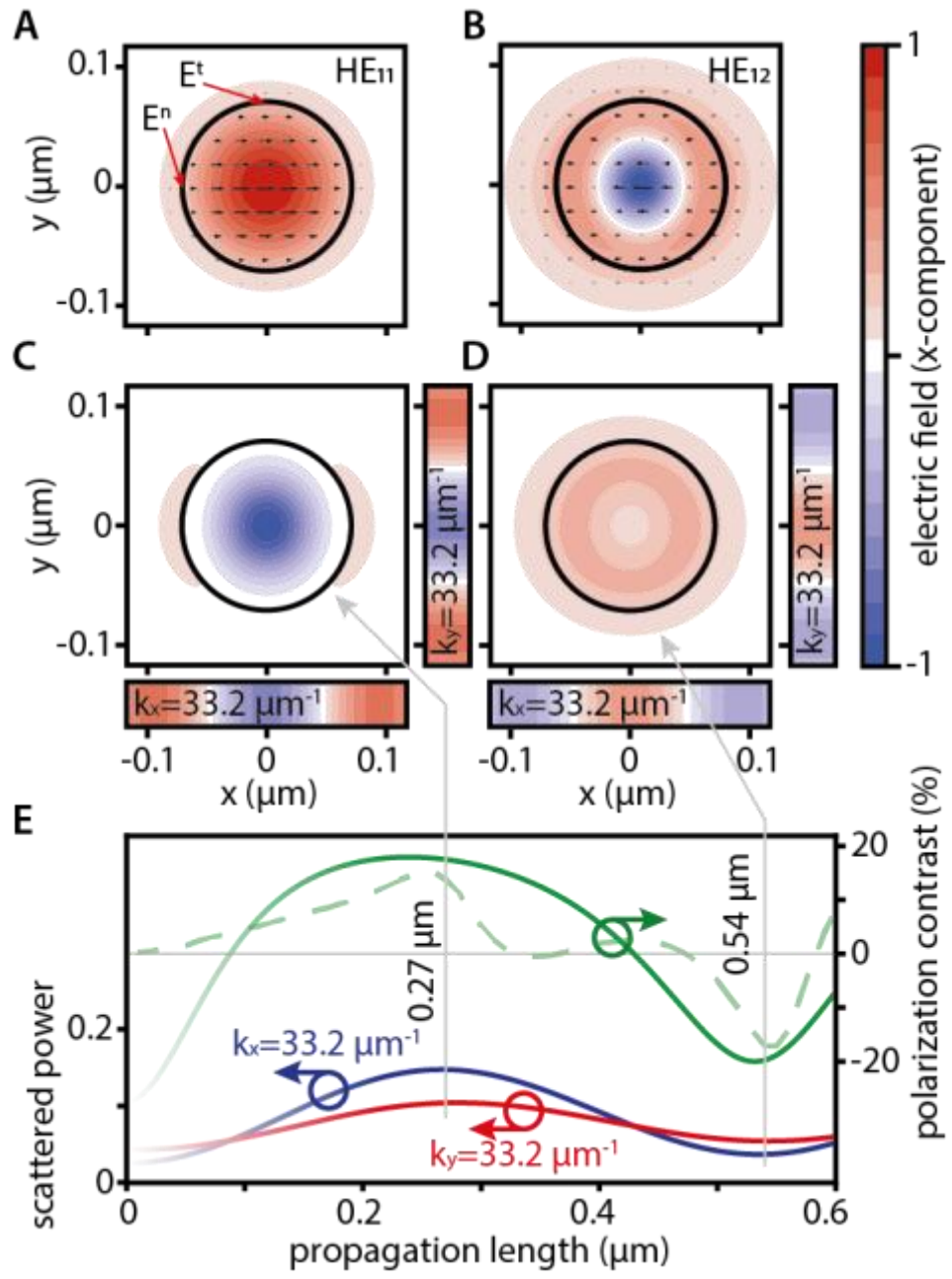

**Fig. 2. EUV eigenmodes.** An x-polarized plane wave incident on a cylindrical void (black circle) in silicon along the cylinder axis can excite **(A)** the $HE_{11}$ and **(B)** the $HE_{12}$ eigenmodes. Arrows denote the modes' transverse electric fields and the false color plots quantify the electric fields' x-components. Spots at which the electric field is aligned primarily tangentially ($E^t$) or perpendicularly ($E^n$) to the vacuum-silicon interface are marked. **(C)** The interfering electric field of both modes after 0.27 μm propagation, where $HE_{11}$ and $HE_{12}$ are out of phase. Horizontal bar: the electric field of a plane wave with $k_x = 33.2\ \mu m^{-1}$. Vertical bar: the electric field of a plane wave with $k_y = 33.2\ \mu m^{-1}$. **(D)** The interfering electric field of both modes after 0.54 μm propagation, where $HE_{11}$ and $HE_{12}$ are in phase (the horizontal and vertical bars are phase-shifted with respect to **(C)** for easier interpretation). **(E)** Propagation-length-dependent scattering efficiency into a plane wave with $k_x = 33.2\ \mu m^{-1}$ (blue line) and into a plane wave with $k_y = 33.2\ \mu m^{-1}$ (red line), i.e., along the dashed lines in Figs. 1C, D. The eigenmode model is not valid before evanescent modes decay, indicated by the white shading. The relative scattering efficiency difference (i.e., the polarization contrast, solid green line) reveals the polarization direction of incident light, an effect that is maintained when integrating the channels in a square lattice (dashed green line, see text).

Whereas the fundamental $HE_{11}$ mode's spatial profile possesses a uniform phase, the $HE_{12}$ mode comprises a central maximum and a π-phase-shifted outer ring separated by a zero crossing. Due to the interface conditions for electromagnetic fields at the channel wall, and because a mode's polarization determines how its electric field aligns with the channel's interfaces (see Fig. 2A), both modes are elongated along their main polarization axis (here, along the x-direction).

Due to their different spatial profiles and overlap with the cladding, different eigenmodes accumulate phase at different rates determined by their effective refractive indices ($n^{\mathrm{HE}_{11}} = 0.982$, $n^{\mathrm{HE}_{12}} = 0.905$). Thus, the simultaneously excited eigenmodes beat upon propagation, spatially redistributing light along the channel with a periodicity given by the excited modes' refractive index difference $z_{\mathrm{beat}} = \frac{\lambda_0}{n^{\mathrm{HE}_{11}} - n^{\mathrm{HE}_{12}}}$ (see supplementary materials). A maximal difference is expected between summing out-of-phase and in-phase modes, i.e., at propagation lengths $z_{\mathrm{beat}}/2 = 0.27$ µm and $z_{\mathrm{beat}} = 0.54$ µm.
Figs. 2C, D show the field profiles at these propagation lengths: After propagating for $z_{\mathrm{beat}}/2 = 0.27$ µm, the interfering modes produce a central maximum and two out-of-phase lobes along the x-direction. If the waveguide ends here, the lobes increase the field profile's overlap with plane waves with transverse momenta $k_x = \pm 33.2$ µm$^{-1}$ and decrease its overlap with plane waves with transverse momenta $k_y = \pm 33.2$ µm$^{-1}$ (see bars in Fig. 2C), i.e., light scatters preferentially in the TM-plane. In accordance with full-field modeling (Figs. 1C, D), the situation reverses for $z_{\mathrm{beat}} = 0.54$ µm (Fig. 2D).

Fig. 2E displays the propagation-length-dependent scattered powers $P_x$ (in the $k_x = 33.2$ µm$^{-1}$ direction) and $P_y$ (in the $k_y = 33.2$ µm$^{-1}$ direction) predicted by the model: both powers peak near 0.27 µm propagation length, but light preferentially scatters into the TM-plane. To quantify this effect in a single observable, we use the polarization contrast

$$\mathrm{PC}(\theta) = \frac{P_x(\theta) - P_y(\theta)}{P_x(\theta) + P_y(\theta)} = \mathrm{PC}_{\mathrm{max}}(1 - 2\sin^2(\theta)) \quad (1)$$

which measures the difference between the scattered powers, depends on the incident polarization direction θ, and is independent of the incident power. Fig. 2E also shows the propagation-length-dependent polarization contrast for x-polarized incident light and predicts a maximum polarization contrast $\mathrm{PC}_{\mathrm{max}} = +18$ % at $z_{\mathrm{beat}}/2 = 0.27$ µm, i.e., up to 44 % more light can be scattered into the TM-plane than into the TE-plane. For unpolarized incident light, the rotational symmetry around the optical axis enforces equal scattering powers $P_x = P_y$ and thus a vanishing polarization contrast.

EUV polarization analyzer: To experimentally demonstrate that EUV multimode photonics enables polarization-dependent light shaping, we implement a nanophotonic polarization analyzer that reveals the photon-energy-resolved polarization of light in a single measurement. We switch from a fully encapsulated void to a square lattice (grating) of cylindrical channels etched into a 0.5-µm-thick crystalline silicon membrane. Fig. 3 illustrates the device's schematic. When considering the inverted roles of material and vacuum, our design resembles recent nanograting-based polarization analyzers for visible and near-infrared light, which achieve a similar function by other working mechanisms (*34*–*36*). To add up light scattered by each channel constructively, we match the lattice momentum $k_\Lambda = 2\pi / \Lambda = 33.2\ \mu m^{-1}$, and thus the propagation direction of the first diffraction orders, to the direction of maximum polarization contrast (yielding a lattice constant $\Lambda = 0.19\ \mu m$). In the following, we address diffraction orders by their transverse momentum vector in units of the lattice momentum, i.e., (-1, 0) corresponds to the order with $k_x = -k_\Lambda$ and $k_y = 0$. The polarization contrast after the grating can be calculated using its order-resolved diffracted powers $P_{(\pm 1,0)}, P_{(0,\pm 1)}$ in Eq. 1.

To verify that the ensemble maintains the polarization-analysis capabilities of the single void, we modeled the lattice's performance using rigorous coupled-wave analysis (green dashed line in Fig. 2E). The simulations reproduce the polarization contrast behavior predicted by the single-particle eigenmode model and suggest that a channel depth of 0.26 μm can yield a maximum polarization contrast of 15 % at 30 eV photon energy. Fig. S3 explores the geometry parameter space and photon-energy-dependent performance and confirms an optimized design that can provide polarization analysis throughout a broad energy range between 28 eV and 50 eV.

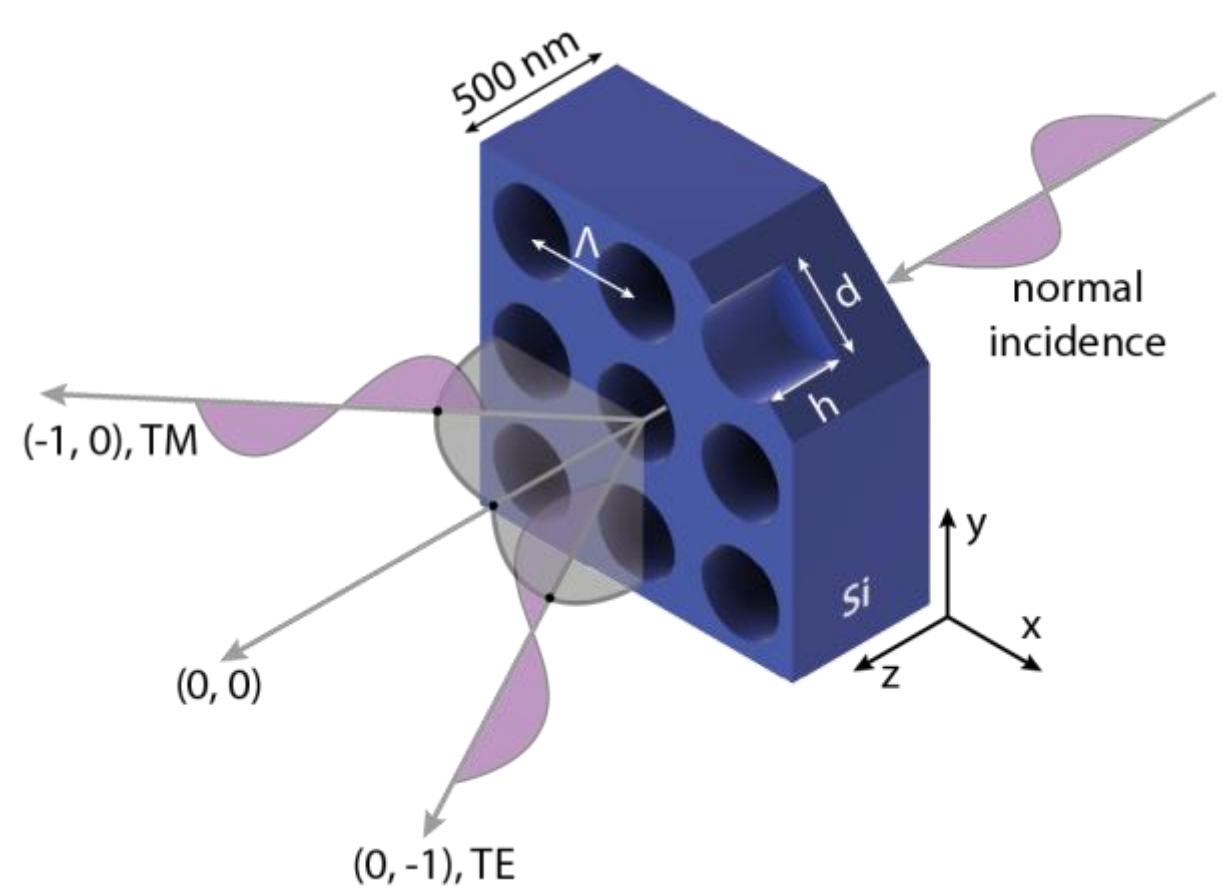

**Fig. 3. EUV polarization analyzer.** Normally incident x-polarized EUV light (purple) transmits through a silicon membrane (blue) with a square array of cylindrical channels (periodicity along the x and y directions Λ, diameter d, etch depth h). The channels scatter light with a direction-dependent efficiency defined by the incoming polarization, and their periodic arrangement diffracts light into discrete diffraction orders. For clarity, the figure only shows the (-1, 0) grating order (in the transverse magnetic plane, TM) and the (0, -1) grating order (in the transverse electric plane, TE).

Following the design (d = 0.14 μm, Λ = 0.19 μm, h = 0.26 μm), we fabricated a 100 μm × 100 μm-large device via electron-beam lithography and inductively coupled plasma reactive ion etching of a 0.5-μm-thick crystalline silicon membrane (see supplementary materials). Figs. 4A, B present scanning electron micrographs of the fabricated polarization analyzer.

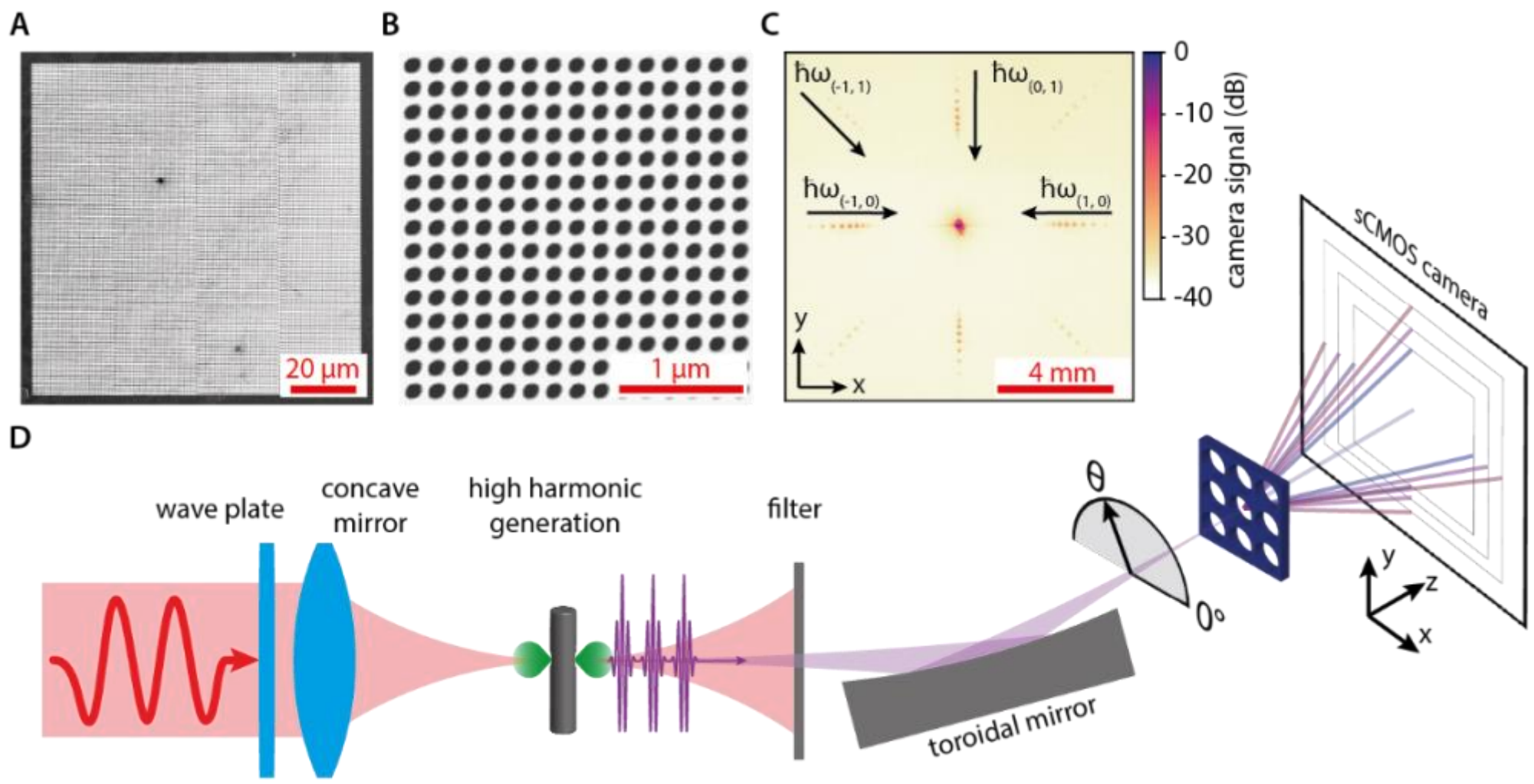

**Fig. 4. Experimental characterization.** (**A**) Scanning electron micrograph of the fabricated 100 μm × 100 μm-large polarization analyzer. Individual write fields are discernible. (**B**) Zoomed-in scanning electron micrograph. (**C**) Two-dimensional diffraction pattern generated by the polarization analyzer showing the zeroth, first, and higher diffraction orders. In addition to polarization analysis, the grating spectrally separates incoming light, revealing the photon-energy-resolved polarization (photon energy: ħω). (**D**) Experimental setup: a half-wave plate controls the linear polarization direction of near-infrared light before it generates EUV high-harmonic radiation that inherits its polarization. After filtering, a grazing-incidence toroidal mirror focuses the EUV light on the polarization analyzer, and a camera detects its diffraction pattern.

<u>Experimental characterization:</u> To investigate the polarization analyzer, we generated an EUV attosecond pulse train via high-harmonic generation by focusing a near-infrared laser pulse (central vacuum wavelength 1.035 μm, pulse duration <190 fs, pulse energy 1 mJ) into an argon gas target. An aluminum filter foil blocked residual near-infrared radiation before a toroidal mirror imaged the transmitted EUV radiation. The polarization analyzer was placed within the Rayleigh range of the focus and 15.6 mm before an EUV-sensitive camera (see Fig. 4D). Turning a half-wave plate upstream of the gas target set the linear polarization direction of the driving near-infrared laser, and consequently of the attosecond pulse train.

Fig. 4C shows a camera image of the polarization analyzer's diffraction pattern containing the zeroth ((0, 0), undeviated beam), first (±1, 0), (0, ±1), and higher diffraction orders. The power in all generated first diffraction orders (averaged over all observed photon energies) makes up ~12 % (~3 % per order) of the power transmitted through an unpatterned portion of the silicon membrane, whose complex refractive index predicts an average transmission of 14 %. Fabricating the analyzer from a 0.26-μm-thick membrane would increase the transmission to 35 %.

Realized polarization contrast: Fig. 5A compares the measured photon-energy- and diffraction-order-resolved EUV power after the polarization analyzer for linearly polarized incident light oriented along the x-direction ($\theta$=0°, parallel to one grating direction, see Fig. 4D); the data evaluation is detailed in the supplementary materials. All first diffraction orders resolve the high-harmonic radiation's spectrum with peaks at odd harmonics of the driving laser frequency.

As expected, the polarization analyzer diffracts more power into the (±1, 0) orders (TM-plane) than into the (0, ±1) orders (TE-plane) at all detected photon energies. Rotating the EUV light's polarization direction $\theta$ interchanges the assignment between the diffraction orders and the TM- and TE-planes. Accordingly, we observe a positive polarization contrast at $\theta$=0°, 180° and 360° and a negative polarization contrast at $\theta$=90° and 270°. Fig. 5B shows the rotation-dependent polarization contrast detected for 30 eV photon energy and a fit of Eq. 1 to the data. The fit yields a polarization contrast amplitude $\mathrm{PC}_{\mathrm{max}}$ = (9.6 ± 0.3) % (95 % confidence), i.e., the polarization analyzer diffracts ~21 % more light into the diffraction orders in the TM-plane compared to the orders in the TE-plane. Fig. 5C presents the simultaneously measured photon-energy-dependent $\mathrm{PC}_{\mathrm{max}}$. In accordance with numerical modeling, the polarization contrast decreases towards higher photon energies. In our experiments, the analyzer can measure the polarization direction of light beyond 44 eV photon energy (28 nm vacuum wavelength), which results in a working bandwidth sufficing for characterizing light pulses lasting shorter than 0.25 fs. Tab. S2 summarizes quantitative results and uncertainties for different photon energies.

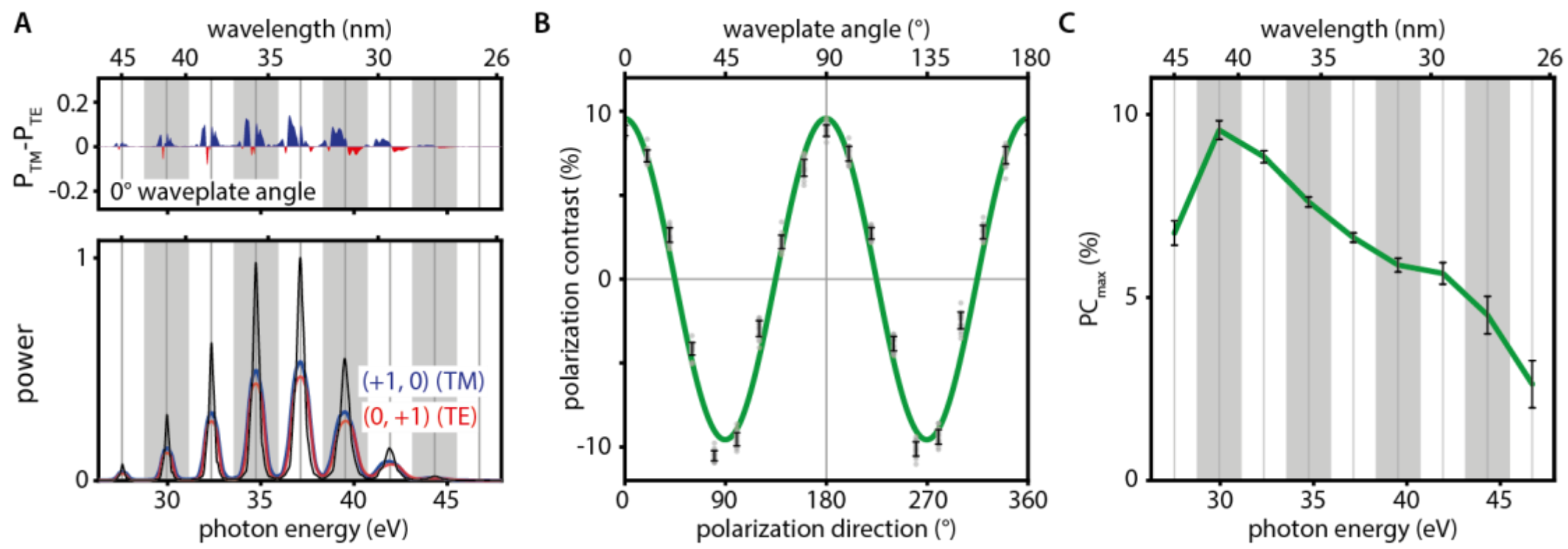

**Fig. 5. Polarization-dependent diffraction and realized polarization contrast of the analyzer. (A)** For x-polarized incident light, the diffracted power in the (+1, 0) diffraction order (which lies in the transverse magnetic (TM) plane, blue line) exceeds the power in the (0, +1) diffraction order (in the transverse electric (TE) plane, red line) for all photon energies. Both spectra were smoothed by an integral-preserving moving average to show their diffracted-power-difference. The black line shows their unsmoothed average. The upper panel shows the unsmoothed photon-energy-resolved difference of the diffracted powers. The gray lines and shaded areas mark expected odd harmonic photon energies. **(B)** Turning a waveplate before high-harmonic generation rotates the EUV light polarization by twice the waveplate rotation angle. Gray dots represent individual measurements (1 second integration time) of the detected polarization contrast at 30 eV photon energy. Black error bars denote the standard deviation of 20 measurements. A fit of Eq. 1 (green line) matches the data well and reveals the polarization contrast amplitude. **(C)** The photon-energy-dependent polarization contrast amplitude ($PC_{max}$, green line) determined via fitting the polarization contrast of each high-harmonic order. Black error bars denote 95 % confidence intervals.

The demonstration uses a square waveguide array, which, by symmetry, is the best choice for retrieving the polarization direction and spectrum of linearly polarized incident light. Arranging the same waveguides in a hexagonal lattice adds a third, independent diffraction order pair, which simultaneously reveals the degree of linear polarization. For fully polarized sources, e.g., high-harmonic generation and many undulators (*37*), the degree of circular polarization follows, and a single measurement using a hexagonal polarization analyzer locates the incident polarization state on the Poincaré sphere except for whether the state lies in the northern or southern hemisphere (see supplementary materials).

Because our polarization analyzer measures the orientation of the polarization axis via a simultaneous, differential measurement, it is, given sufficient brightness, capable of single-shot operation, a crucial capability for characterizing sources with fluctuating or unrepeatable power, spectrum, and temporal structure, e.g., free-electron lasers.

As with common polarization analyzers, the polarization contrast our device generates changes most steeply with incident polarization-direction variations when the power diffracted into the TM- and TE-planes is balanced, i.e., around $\theta = 45°$. The standard deviations in Fig. 5B suggest that, set to this angle, the polarization analyzer resolves polarization rotations as small as $\Delta\theta = 0.9°$ in a 1-second measurement of our high-harmonic-generation source, sufficient to observe reported Faraday rotations (*38*, *39*) and magneto-optic Kerr rotations (*40*) of magnetic thin films. Thus, tailoring our analyzer design to match the characteristic absorption edges of transition metals promises to facilitate studying magnetization and spin-transfer dynamics with attosecond temporal resolution and element specificity (*1*–*6*).

The realized polarization contrast amounts to ~64 % of the numerical prediction despite the etched channels' roughness, uncertainties in the achieved etch depth (see Fig. S4), a possible oxidation of the analyzer surface, uncertainties in silicon's refractive index, and a possible ellipticity of the incident polarization states (see supplementary materials). Our device demonstrates that multimode photonics is a viable approach for implementing transmissive optics for EUV light, a spectral region where material absorption has so far prevented realizing many crucial optical functions. Beyond polarization analysis and control, the concept generates new opportunities for nanoscopic light shaping in EUV metaoptics. For 41-nm-wavelength light, incident plane waves only excite the $HE_{11}$ mode in channel waveguides with diameters smaller than 100 nm, which are already exploited to design metasurfaces that shape the spatial phase of transmitted EUV light (*22*, *23*). In waveguides with a larger diameter, the $HE_{12}$ mode is also excited, opening mode beating as a design parameter and allowing intra-unit-cell phase shaping (see Fig. 2), i.e., metasurfaces with spatial resolutions finer than the fabrication limit. Close to a mode's cutoff frequency, its effective refractive index becomes considerably frequency dependent. Exploiting this fact with tailored waveguides could therefore bring slow-EUV-light (*41*)-enhanced light-matter interaction, and attosecond pulse dispersion shaping, within reach of transmissive nanophotonics.

**Acknowledgements:** We acknowledge Joachim R. Krenn and Andreas Hohenau from the University of Graz for providing access to and support with their electron beam lithography system.

**Funding:** This research was supported by the Scientific Service Units of ISTA through resources provided by the Nanofabrication Facility. M.O. acknowledges funding from the European Union (grant agreement 101076933 EUVORAM). The views and opinions expressed are, however, those of the author(s) only and do not necessarily reflect those of the European Union or the European Research Council Executive Agency. Neither the European Union nor the granting authority can be held responsible for them.

**Author Contributions:** M.O. conducted the numerical modeling and designed the polarization analyzer. A.K., L.W. and D.J.G. developed and carried out the lithography process. A.K. and R.P. etched and imaged the samples. A.K., A.S.G., M.S., and M.O. designed and set up the laser frontend and EUV beamline. A.K., L.W., D.J.G. and M.O. conducted the experiment and analyzed the experimental data. M.O. wrote the manuscript with input from A.K. and M.S. All authors discussed the final version of the manuscript. M.O. acquired the project funding.

**Competing interests:** The authors declare no competing interests.

# Supplementary Materials for

## Nanophotonic Extreme Ultraviolet Polarimeter

Anna Kulter[1], Lorenz Weiß[1], David J. Grafinger[1], Alexander Sagar Grossek[1], Rodolfo Previdi[2], Martin Schultze[1], Marcus Ossiander[1, 3, *]

1 Institute of Experimental Physics, Graz University of Technology, Austria.
2 Institute of Science and Technology Austria, Austria.
3 John A. Paulson School of Engineering and Applied Sciences, Harvard University, USA.
* Corresponding author: marcus.ossiander@tugraz.at

Numerical Modeling: We used the published complex refractive index of silicon (*42–44*) for all simulations. Scattering simulations: We modeled light scattering from isolated vacuum cylinders in silicon using finite-difference-time-domain simulations (Flexcompute Inc., tidy3d), a 0.5 μm × 0.5 μm × 1.1 μm-large simulation volume, a 2 nm discretization, an x-polarized total-field scattered-field source, and perfectly matched layers along all simulation boundaries. Eigenmodes: We calculated the eigenmodes of an isolated cylindrical void in silicon using a two-dimensional frequency domain eigenmode solver (Flexcompute Inc., tidy3d), a 0.6 μm × 0.6 μm-large simulation area, a 2 nm discretization, and perfectly matched layers along the transverse boundaries of the area. Gratings: We calculated the grating's diffraction efficiencies and polarization contrast using rigorous coupled-wave simulations (GRCWA (*45*)). The unit cell size corresponded to the grating periodicity, and the simulation contained a single channel etched into the exit facet of the grating. The structure was excited with an x-polarized normally incident plane wave. The implicit boundaries were periodic along the transverse directions and radiating along the optical axis.

EUV eigenmodes: A vacuum void encapsulated in silicon possesses a refractive index contrast of $n_{vacuum} - n_{Si} = 0.14$, indicating strong guiding (*33*) and the necessity to investigate its vectorial (hybrid) modes. A numerical aperture $NA = \sqrt{(n_{vacuum}^2 - n_{Si}^2)} = 0.51$ , a vacuum wavelength $\lambda_0 = 41.3$ nm, and a diameter of 0.14 μm yield a normalized frequency $V = \pi\, d\, NA / \lambda_0 = 5.43$ which suggests 16 supported eigenmodes (see Tab. S1, Fig. S1, Fig. S2). A normally incident plane wave only excites the $HE_{11}$ and $HE_{12}$ modes.

At the cylinder wall (the material interface between vacuum and silicon), the tangential component of the modes' electric field $E^t_{\text{vacuum}} = E^t_{\text{Si}}$ is continuous, and its normal component $\varepsilon_{\text{vacuum}} E^n_{\text{vacuum}} = \varepsilon_{\text{Si}} E^n_{\text{Si}}$ changes abruptly according to the permittivity ratio $\varepsilon_{\text{vacuum}}/\varepsilon_{\text{Si}}$, elongating both modes along their main polarization axis. When neglecting evanescent modes, the resulting propagation-length-z-dependent electric field profile $\boldsymbol{E}(x,y,z) = 0.70\boldsymbol{E}^{\text{HE}_{11}}(x,y,0)e^{ik_0(n^{\text{HE}_{11}}+i\kappa^{\text{HE}_{11}})z} + 0.53\,\boldsymbol{E}^{\text{HE}_{12}}(x,y,0)e^{ik_0(n^{\text{HE}_{12}}+i\kappa^{\text{HE}_{12}})z}$ along the vacuum cylinder can be described using the transverse electric field profiles $\boldsymbol{E}^{\text{HE}_{11}}$ and $\boldsymbol{E}^{\text{HE}_{12}}$, and the modes' complex effective refractive indices $n^{\text{HE}_{11}} + i\kappa^{\text{HE}_{11}}$ and $n^{\text{HE}_{12}} + i\kappa^{\text{HE}_{12}}$ (see Tab. S1). This yields a beating periodicity given by the modes' refractive index difference $z_{\text{beat}} = \frac{\lambda_0}{n^{\text{HE}_{11}} - n^{\text{HE}_{12}}}$.

<u>Fabrication:</u> The device was fabricated by spin-coating a positive electron-beam resist (Allresist GmbH, AR-P 6200.13, CSAR 62) onto a 0.5-µm-thick crystalline silicon membrane (Silson Ltd, M1003577). The hole pattern was exposed in the 0.4-µm-thick resist using electron-beam lithography (Raith GmbH, eLine plus) and subsequently transferred into the silicon via pseudo-Bosch inductively coupled plasma reactive ion etching (ICP-RIE, Oxford Instruments, Plasma Pro 100 Cobra, etching gas: $SF_6$, passivation gas: $C_4F_8$) for 160 seconds. The pseudo-Bosch process etches and passivates simultaneously, suppressing scalloping and resulting in smoother channel sidewalls than the standard Bosch process. Remaining resist was removed with oxygen plasma ashing.

<u>Input polarization states:</u> High-harmonic generation in an isotropic medium maintains a linearly polarized driver's polarization upon conversion. In our experiment, a half-wave plate turned by $\theta/2$ rotated the driving laser and the resulting EUV radiation to polarization states with the Jones vectors $|\theta\rangle = \begin{pmatrix}\cos(\theta)\\ \sin(\theta)\end{pmatrix}$. The reflection phase of the boron-carbide-coated grazing-incidence toroidal mirror (86° incidence angle) is polarization-dependent (*42*). Thus, it acts as a waveplate with the Jones matrix $T = \begin{pmatrix}0.94\,e^{3^\circ i} & 0\\ 0 & 0.97\,e^{-171^\circ i}\end{pmatrix}$. While this does not affect polarizations aligned with the x- or y-axis in Fig. 4, the toroidal mirror swaps the $|-45^\circ\rangle$ and $|+45^\circ\rangle$ polarization states and introduces a very slight ellipticity: $T|-45^\circ\rangle = \begin{pmatrix}0.66\,e^{3^\circ i}\\ 0.68\,e^{9^\circ i}\end{pmatrix} \approx 0.96|+45^\circ\rangle$. As the beam is predicted to still be >99 % linearly polarized, we treat it as linearly polarized in the rest of the paper.

Experimental data evaluation: We calibrated the absolute photon energy scale of our polarization analyzer using the characteristic absorption of germanium at 32 eV photon energy, see Fig. S5. For each incident polarization state, we recorded the polarization analyzer's diffraction pattern using a cooled, back-thinned, back-illuminated scientific complementary metal-oxide-semiconductor (sCMOS) camera (Excelitas Technologies Corp., pco.edge 4.2 bi XU) and an integration time of 1 second. After subtracting a constant background, we integrated over a 20-pixel-wide line along each diffraction order to calculate the energy-resolved diffracted power. We then summed the detected power in a 2.4-eV-wide window around each high-harmonic order to calculate the polarization contrast. To determine its uncertainty, we recorded 20 pictures for each incident polarization state and independently evaluated the polarization contrast from each picture. Lastly, we fitted Eq. 1 to the polarization-state-dependent data for each high-harmonic order to find the polarization contrast amplitude. Its 95 % confidence interval was calculated using the fit's covariance matrix and the two-sided quantiles of the Student's t-distribution.

Hexagonal polarization analyzer: Switching from a square waveguide array with lattice vectors $\begin{pmatrix}1\\0\end{pmatrix}\Lambda$ and $\begin{pmatrix}0\\1\end{pmatrix}\Lambda$ to a hexagonal waveguide array with lattice vectors $\begin{pmatrix}0\\1\end{pmatrix}\Lambda$ and $\begin{pmatrix}\sqrt{3}\\1\end{pmatrix}\frac{\Lambda}{2}$ yields a polarization analyzer that generates three independent diffraction order pairs (±1, 0), (0, ±1), (+1, +1)/(-1, -1) carrying polarization information. The polarization contrast amplitude of the hexagonal analyzer can be calibrated using the diffracted powers $P_{(1,0)}, P_{(0,1)}, P_{(1,1)}$ for normally incident light that is linearly polarized along any direction and $\mathrm{PC_{max}} = \frac{2\sqrt{P_{(0,1)}^2+P_{(1,1)}^2+P_{(1,0)}^2-P_{(0,1)}P_{(1,1)}-P_{(1,1)}P_{(1,0)}-P_{(1,0)}P_{(0,1)}}}{P_{(0,1)}+P_{(1,1)}+P_{(1,0)}}$. The third diffraction order's additional information then allows calculating the dominant polarization direction
$\theta = \frac{1}{2}\tan^{-1}\left(\frac{\mathrm{M}}{N}\right)$ and the degree of linear polarization $\mathrm{DoLP} = \sqrt{N^2 + M^2}$ of any incident polarization state using the helper quantities $T = P_{(0,1)} + P_{(1,1)} + P_{(1,0)}$, $N = \left(2\,P_{(0,1)} - P_{(1,1)} - P_{(1,0)}\right)/(T\,\mathrm{PC_{max}})$, and $M = \sqrt{3}\,\left(P_{(1,1)} - P_{(1,0)}\right)/(T\,\mathrm{PC_{max}})$. If the observed source is known to be fully polarized, the degree of circular polarization $\mathrm{DoCP} = \sqrt{1 - N^2 - M^2}$ follows and reveals the magnitude of the phase difference $|\delta| = \cos^{-1}\left(\frac{\mathrm{M}}{\sqrt{1-\mathrm{N}^2}}\right)$ between the x- and y-components of the incident EUV light.

| mode | $V_{cutoff}$ (normalized frequency cutoff (*46*)) | $n_{eff}$ | $\kappa_{eff}$ | overlap (normalized) with a normally incident x-polarized plane wave |
|---|---|---|---|---|
| $HE_{11}$ | 0.0 | 0.98 | 3.6e-4 | 1 |
| $HE_{11}$ | 0.0 | 0.98 | 3.6e-4 | - |
| $TE_{01}$ | 2.405 | 0.96 | 8.3e-4 | - |
| $HE_{21}$ | 2.42 | 0.95 | 1.0e-3 | - |
| $HE_{21}$ | 2.42 | 0.95 | 1.0e-3 | - |
| $TM_{01}$ | 2.405 | 0.95 | 1.2e-3 | - |
| $EH_{11}$ | 3.83 | 0.92 | 2.0e-3 | - |
| $EH_{11}$ | 3.83 | 0.92 | 2.0e-3 | - |
| $HE_{31}$ | 3.86 | 0.92 | 2.2e-3 | - |
| $HE_{31}$ | 3.86 | 0.92 | 2.2e-3 | - |
| $HE_{12}$ | 3.83 | 0.91 | 3.0e-3 | 0.73 |
| $HE_{12}$ | 3.83 | 0.91 | 3.0e-3 | - |
| $EH_{21}$ | 5.14 | 0.88 | 3.9e-3 | - |
| $EH_{21}$ | 5.14 | 0.88 | 3.9e-3 | - |
| $HE_{41}$ | 5.16 | 0.87 | 4.5e-3 | - |
| $HE_{41}$ | 5.16 | 0.87 | 4.5e-3 | - |

**Table S1. Eigenmodes of a 0.14-μm-diameter channel in silicon for light with 41.3 nm vacuum wavelength.**

| photon energy (eV) | $PC_{max}$ ± 95 % confidence intervals |
|---|---|
| 27.6 | 6.8 % ± 0.3 % |
| 29.9 | 9.6 % ± 0.3 % |
| 32.3 | 8.8 % ± 0.2 % |
| 34.7 | 7.6 % ± 0.1 % |
| 37.1 | 6.6 % ± 0.1 % |
| 39.5 | 5.9 % ± 0.2 % |
| 41.9 | 5.7 % ± 0.3 % |
| 44.3 | 4.5 % ± 0.5 % |
| 46.7 | 2.6 % ± 0.6 % |

**Table S2. Measured photon-energy-resolved polarization contrast amplitudes.**

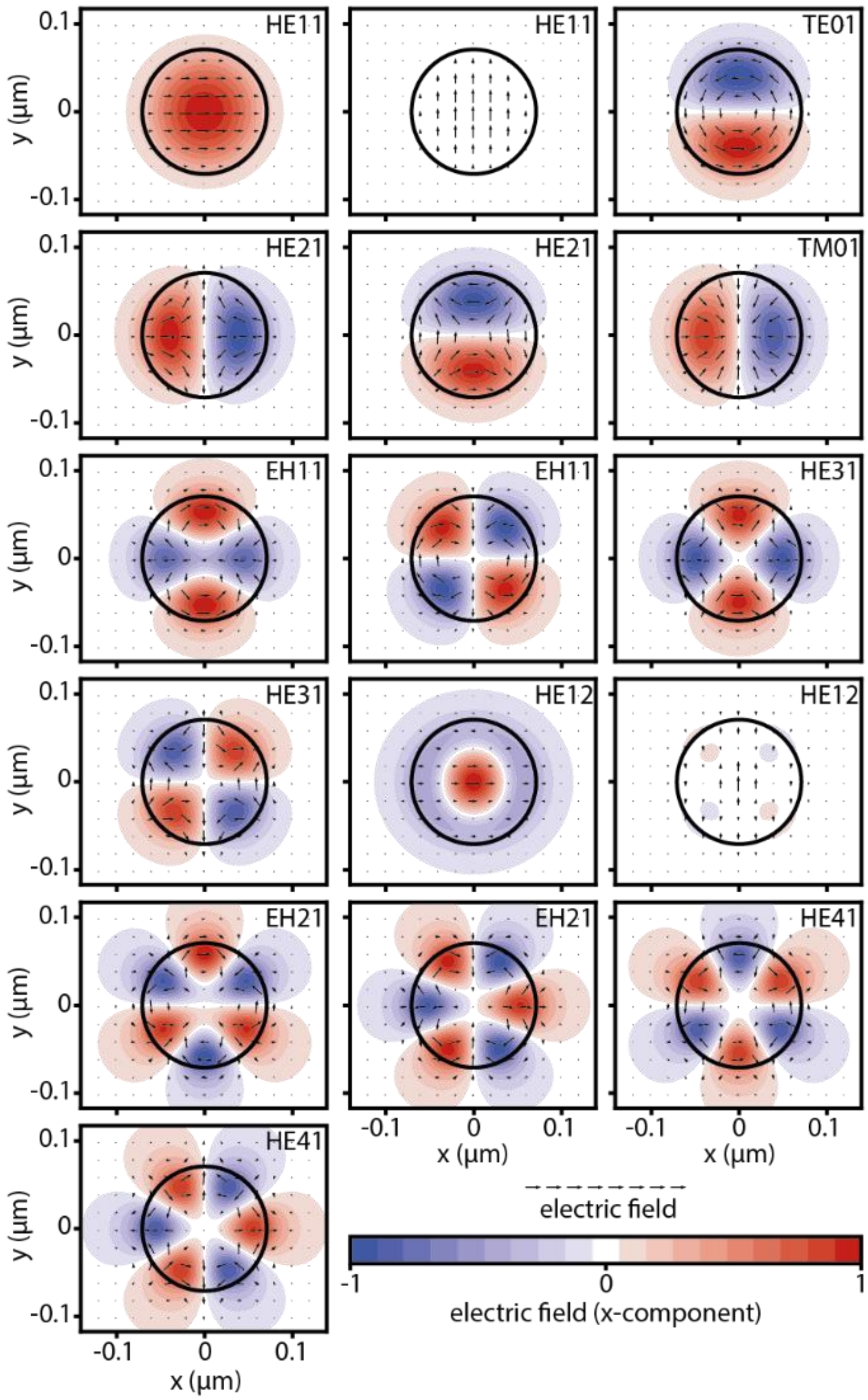

**Fig. S1. Simulated transverse electric field profiles of the propagating eigenmodes of a 0.14-µm-diameter vacuum channel in silicon for light with 41.3 nm vacuum wavelength.** Arrows denote the electric fields' x- and y-components; the false color plots show the electric fields' x-components to allow estimating the coupling to a normally incident x-polarized plane wave.

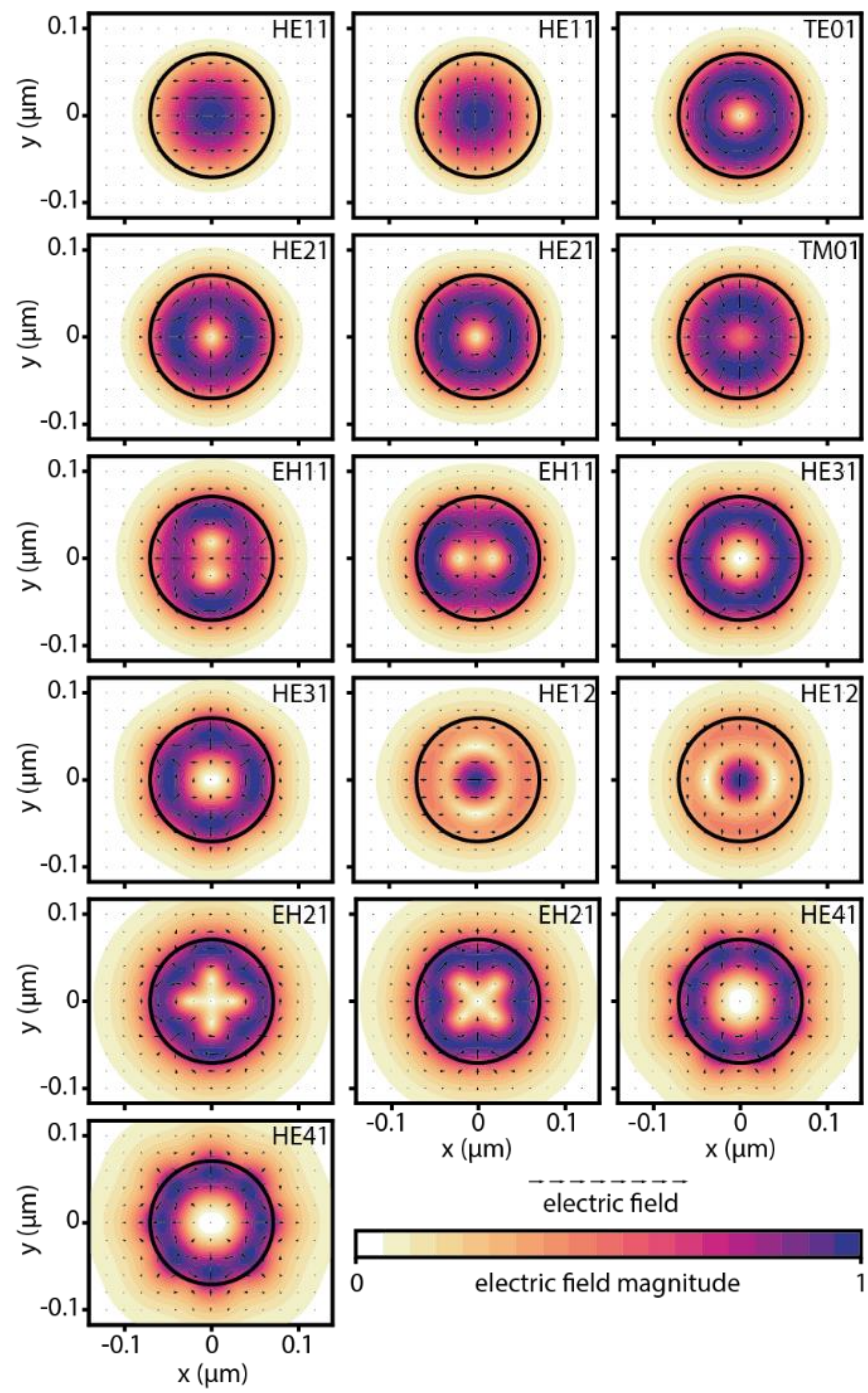

**Fig. S2. Simulated electric field profiles of the propagating eigenmodes of a 0.14-µm-diameter vacuum channel in silicon for light with 41.3 nm vacuum wavelength.** Arrows denote the electric fields' x- and y-components; the false color plots show the electric fields' magnitudes.

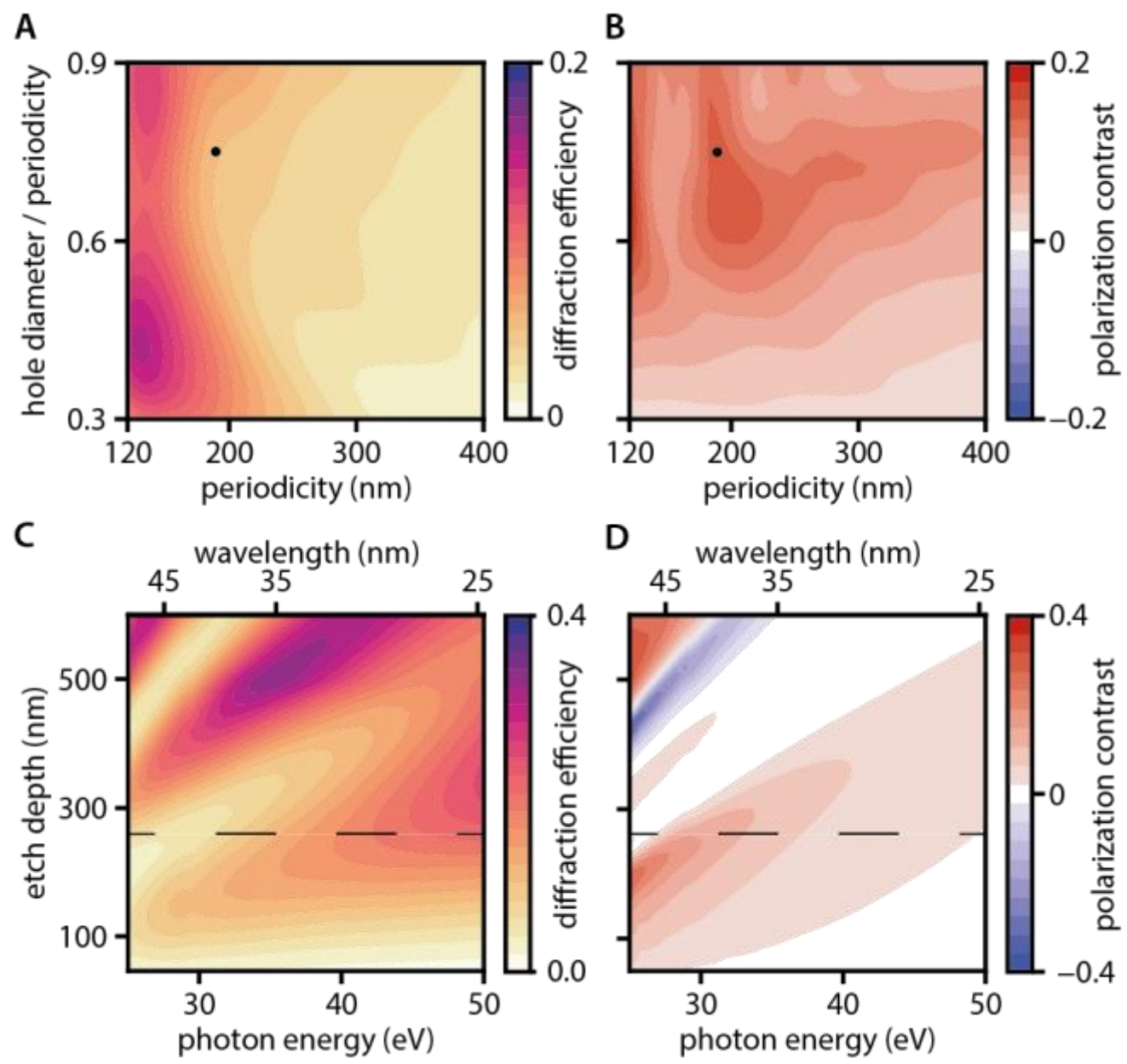

**Fig. S3. Polarization analyzer design via rigorous coupled-wave analysis.** (**A**) Periodicity- and channel diameter-dependent summed diffraction efficiency into the four first diffraction orders of a 0.26-μm-deep (the fabricated depth) channel array etched into a 0.5-μm-thick crystalline silicon membrane. (**B**) Corresponding polarization contrast of the first diffraction orders. The black dots mark the fabricated design. (**C**) Energy- and etch-depth-dependent summed diffraction efficiency into the four first diffraction orders generated by an array of 0.14-μm-diameter channels in a square lattice with 0.19-μm periodicity. (**D**) Corresponding polarization contrast of the first diffraction orders. The dashed line marks the final design. In (**C**) and (**D**), the membrane thickness was increased for etch depths exceeding 0.5 μm.

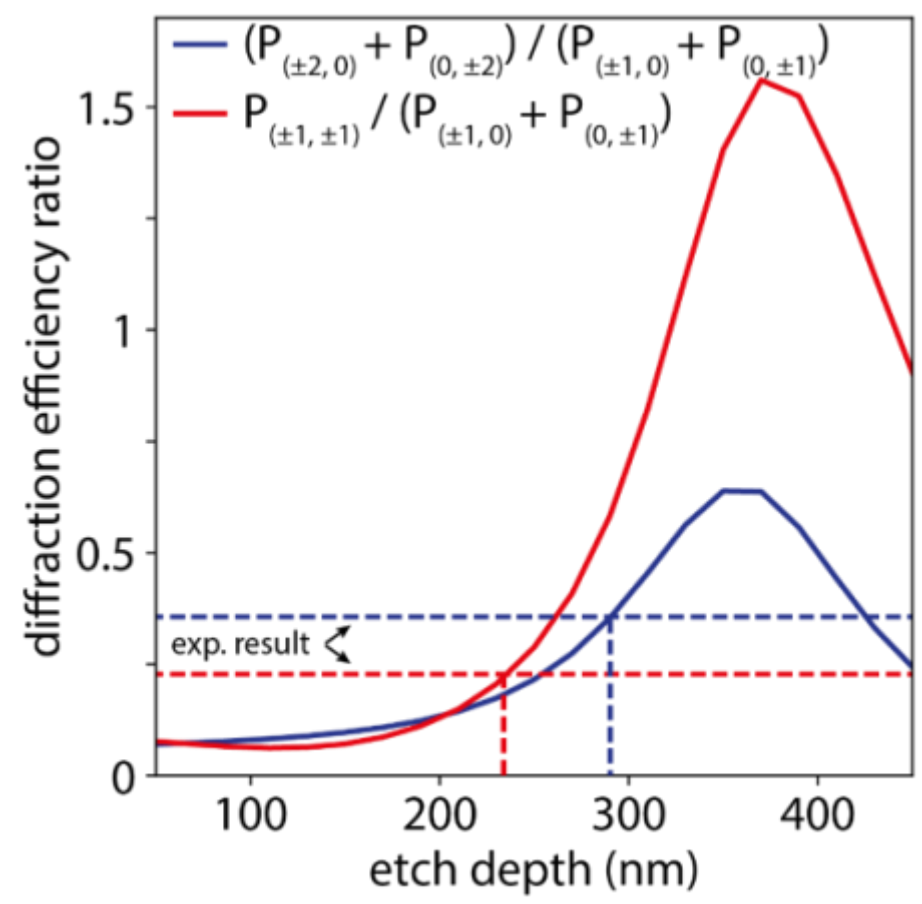

**Fig. S4. Channel depth verification.** Rigorous coupled-wave analysis simulations of the etch-depth-dependent diffraction efficiency ratio for the first, second, and first diagonal diffraction orders (solid lines) compared to the experimentally determined ratios (dashed lines). The experimentally observed ratio of the power in the second $P_{(\pm2,\,0)} + P_{(0,\,\pm2)}$ diffraction orders to the power in the first diffraction orders $P_{(\pm1,\,0)} + P_{(0,\,\pm1)}$ suggests ~0.29-µm deep channels. The experimentally observed ratio of the power in the diagonal $P_{(\pm1,\,\pm1)}$ diffraction orders to the power in the first diffraction orders $P_{(\pm1,\,0)} + P_{(0,\,\pm1)}$ suggests ~0.24-µm deep channels. The average fits our expected 0.26-µm etch depth.

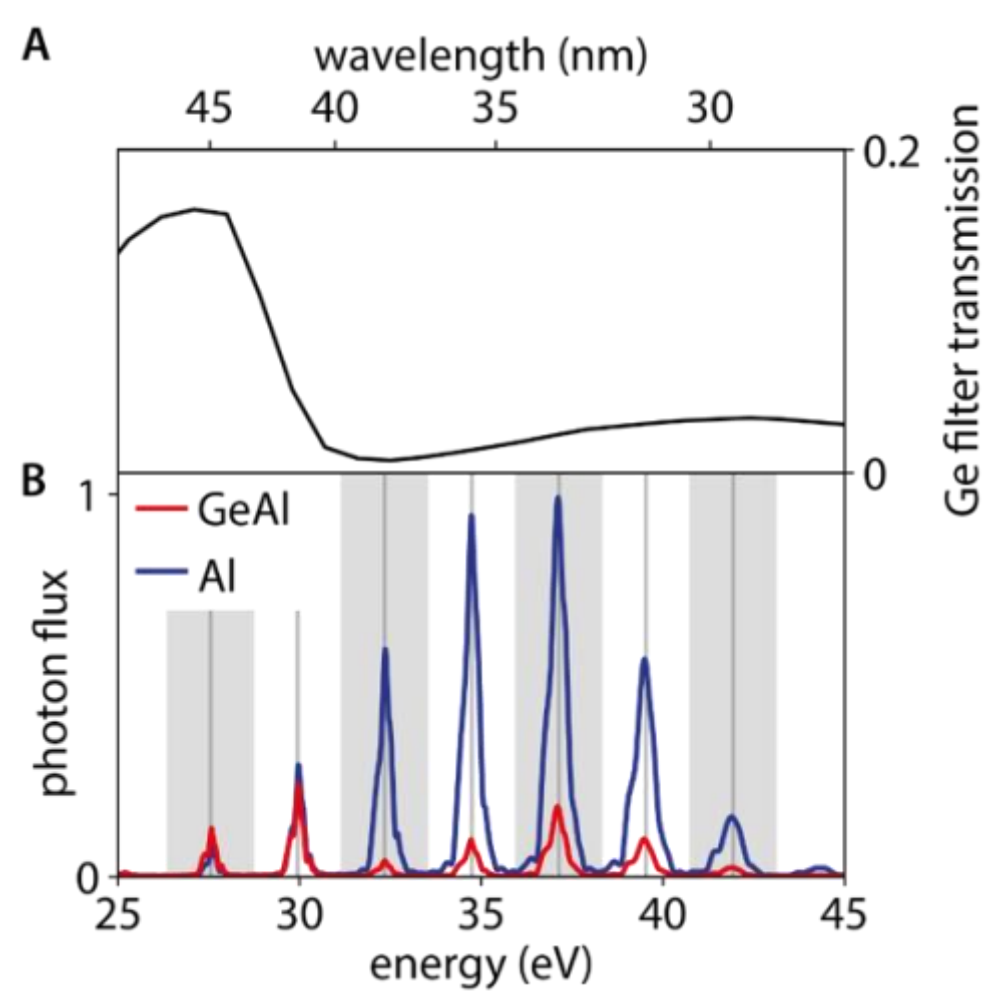

**Fig. S5. Energy calibration using a germanium filter.** (**A**) Literature energy-dependent filter transmission ***(43)***. (**B**) Comparison of the energy-dependent photon flux detected after a 0.50-µm-thick aluminum filter (blue) and after a combined 0.17-µm germanium and 0.20-µm aluminum filter (red). Germanium's transmission minimum at 32 eV determines the absolute energy position of the observed high-harmonic emission maxima. With this anchor, we fine-tune the analyzer-detector distance to move the harmonics to odd multiples of the driving laser photon energy. An analyzer-detector distance of 15.6 mm reproduces the germanium transmission function, the high-harmonic-energy positions, and our measured analyzer-detector distance well.